# Harmonic generation from metal-oxide and metal-metal boundaries

M. Scalora[1*], M.A. Vincenti[2], D. de Ceglia[3], N. Akozbek[4], M. J. Bloemer[1], C. De Angelis[2], J. W. Haus[5], R. Vilaseca[6], J. Trull[6], and C. Cojocaru[6]

[1] *Charles M. Bowden Research Center, AMRDEC, RDECOM, Redstone Arsenal, AL 35898-5000 - U.S.A.*

[2] *Department of Information Engineering – University of Brescia, Via Branze 38 25123 Brescia, Italy*

[3] *National Research Council - AMRDEC, Charles M. Bowden Research Center, Redstone Arsenal, AL 35898-5000 U.S.A.*

[4] *AEgis Technologies Inc., 401 Jan Davis Dr. 35806, Huntsville, AL U.S.A.*

[5] *Department of Electro-Optics and Photonics, University of Dayton, Dayton, OH 45469-2951 U.S.A.*

[6] *Physics Department, Universitat Politècnica de Catalunya, Rambla Sant Nebridi 22, Terrassa, Barcelona, Spain*

[michael.scalora.civ@mail.mil](michael.scalora.civ@mail.mil)*

**Abstract**

We explore the outcomes of detailed microscopic models by calculating second- and third-harmonic generation from thin film surfaces with discontinuous free-electron densities. These circumstances can occur in structures consisting of a simple metal mirror, or arrangements composed of either different metals or a metal and a free electron system like a conducting oxide. Using a hydrodynamic approach we highlight the case of a gold mirror, and that of a two-layer system containing indium tin oxide (ITO) and gold. We assume the gold mirror surface is characterized by a free-electron cloud of varying density that spills into the vacuum, which as a result of material dispersion exhibits epsilon-near-zero conditions and local field enhancement at the surface. For a bilayer consisting of a thin ITO and gold film, if the wave is incident from the ITO side the electromagnetic field is presented with a free-electron discontinuity at the ITO/gold interface, and wavelength-dependent, epsilon-near-zero conditions that enhance local fields and conversion efficiencies, and determine the surface's emission properties. We evaluate the relative significance of additional nonlinear sources that arise when a free-electron discontinuity is present, and show that harmonic generation can be sensitive to the density of the screening free-electron cloud, and not its thickness. Our findings also suggest the possibility to control surface harmonic generation through surface charge engineering.

**Introduction**

The study of second harmonic generation (SHG) from surfaces has remained active since the early days of nonlinear optics [1-8]. The list includes nanoscale surface structures that exhibit strong nonlinear chirality [6, 8]. Surface SHG is an ideal non-destructive tool to study surfaces with sensitivity at the sub-nanometer scale. However, our understanding of the surface properties remains uncertain because while different theoretical models may yield similar angular dependence of the generated SH signal, there is disagreement on the magnitude of the predicted SH signal, with results sometimes differing by several orders of magnitude. This apparent, model-dependent inconsistency is symptomatic of a combination of incomplete knowledge of the nanoscale surface composition and structure, and of the relative significance of a number of quantum-based physical phenomena. For example, the effective electron mass is reported to be sensitive to the particular deposition method employed [9]. Through the detection of surface plasmon modes, it has also been shown that a simple metal layer may be denser on the substrate side compared to its air side, leading to a position dependent dielectric function [10] that models fail to take into account, and discrepancies between actual and tabulated values. At the same time, the dielectric constant itself may be a function of both frequency and wave vector via the excitation of nonlocal effects, e.g. electron gas pressure [11].

In addition to questions surrounding deposition processes and surface preparation, there are issues regarding the methods that are used to predict electrodynamic phenomena in nanoscale systems. Over the decades, technological progress has led to a steady miniaturization process that has resulted in structures having features with near-atomic size. On the sub-nanometer scale, the applicability of classical electrodynamics is called into question: the theory is based on a process that turns the rapidly fluctuating microscopic fields found near individual atoms into macroscopic fields averaged over a volume of space that may contain countless atoms, or dipoles. The medium loses its granularity only to be described as a continuum that necessitates the mere application of boundary conditions at interfaces [12]. This simplified picture fails if the macroscopic theory is applied to systems with feature sizes that compare with the size of atoms [13]. This is already the case for typical nanowire and/or nanoparticle systems that are easily fabricated with features so small and so closely spaced that the electronic wave functions spilling outside their respective surfaces may overlap. The electronic wave function diameter of a typical noble metal atom is approximately 3Å, while the electronic cloud forming and shielding a flat, noble metal surface may extend several Ås into free space (Fig. 1).

The study of light interactions with optically thick metal layers below their plasma frequencies, where the dielectric constant is negative, had been limited to the study of reflection due to the large negative dielectric constant and the absence of propagation modes. At the nanometer scale, transmission through thick metal layers and structures that may contain hundreds of nanometers of metal has been shown to be possible by exploiting cavity/interference phenomena that localize the light inside the metal itself [5, 14], as well as surface plasmon excitation where the light is channeled through subwavelength apertures [15].

The linear optical response of metals is almost always modelled using the Drude model, i.e. as a cloud of free electrons with a frequency dependent dielectric constant and fixed boundaries. At optical wavelengths valence electrons contribute to the dielectric constant. The inadequacy of the Drude model is reflected in experimental observations of light scattering from nanoscale systems in the optical range, and so it is enhanced by hydrodynamic models that incorporate nonlocal effects through terms like electron gas pressure [16-21], and surface and bulk nonlinearities [22-29]. Ultimately, the sub-nanometer gap between metals enables quantum tunneling [26-34], which may also easily be incorporated into dynamical, time domain models [28].

The absorption of free electron systems like ITO or cadmium oxide (CdO) is smaller compared to that of noble metals, especially in the range where the real part of the dielectric constant crosses the axis and takes on *near-zero* values. Materials used in the zero crossing region have been dubbed epsilon-near zero (ENZ) materials [35]. If the imaginary contribution to the permittivity also approaches zero then the refractive index also approaches zero. The so-called *zero index materials* would propagate an electromagnetic wave from one side to the other with no phase delay. While there are interesting consequences of the peculiar dispersive linear optical properties of these materials, our interest is aimed at studying novel, low-intensity nonlinear optical phenomena that otherwise are observed only for high, local fields. ENZ materials contribute to enhanced optical harmonic generation and as such their study can shed new light on our fundamental understanding of both linear and nonlinear optical processes of free electron systems [36].

The nonlinear field enhancement mechanism is triggered by the requirement that the longitudinal component of the displacement vector of a TM-polarized field be continuous, which for homogeneous, flat structures is exemplified by the relationship: $\varepsilon_{in}E^z_{in} = \varepsilon_{out}E^z_{out}$. $\varepsilon_{in(out)}$ is the dielectric constant inside (outside) the medium, and $E^z_{in(out)}$ is the corresponding longitudinal component of the electric field amplitude inside (outside) the material. It follows that if $\varepsilon_{in} \Rightarrow 0$, then $E^z_{in} \Rightarrow \infty$. The field enhancement and the observation of nonlinear [37, 38] and nonlocal effects [39] is primarily limited by the imaginary contribution to the dielectric function of the ENZ material. Additionally, nonlocal effects in these materials can play a major role in noble metals, as both field penetration inside the medium and field derivatives are correspondingly more prominent, leading to significant deviations from the predictions of local electromagnetism. Quantum tunneling contributions and nonlocal effects are simultaneously accounted for in structures with sub-nanometer spacing in reference [28].

**The Model**

We illustrate our theoretical approach using a representative example, the seemingly simple problem of a gold film interface as seen from the atomic scale. In Fig. 1 (a) we depict a typical noble metal atom: a nearly-free, s-shell electron that orbits at an approximate distance $r_s \sim 1.5$Å from the nucleus, and d-shell electrons whose orbits extend out approximately $r_d \sim 0.5$Å from the nucleus [40,41]. The rudimentary picture that emerges even from a cursory look at Fig. 1 (b), which schematically represents atoms distributed at and just below the surface of a hypothetical metallic medium, is one of a negatively charged electron cloud that spills outside the ionic surface

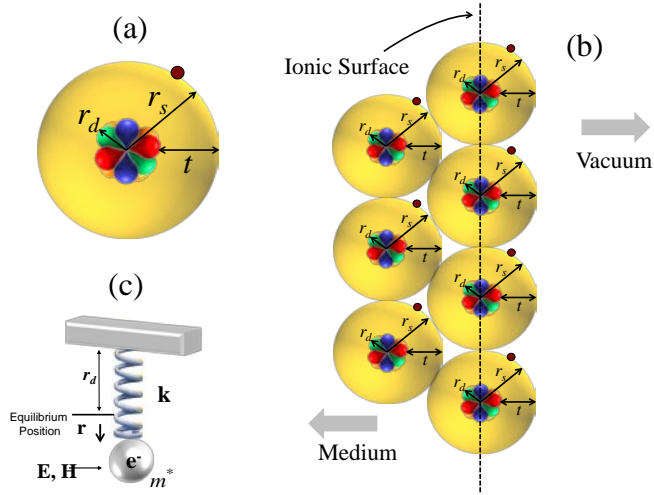

**Figure 1:** (a) Schematic representation of a typical noble metal atom. The radii represent the maximum amplitude of orbital wave functions calculated using a many-body approach. (b) The dashed line represents the last row of atoms

in a medium that extends to the left. This rudimentary picture suggests that a free-electron only patina of approximate thickness *t* shields a medium composed of both free and bound electrons. (c) Diagram of bound, d-shell electrons represented as Lorentz oscillators.

(the dashed line in Fig. 1 (b)) and screens the inside portions of the metal. The figure also suggests that interior sections of the medium contain a combination of free and bound charges, which present their own surface to the incoming electromagnetic wave. Therefore, it is plausible to assume that some of the reasons for the discrepancies between experimental results and most theoretical models, and between theoretical models themselves, may to some extent reside in the failure to accurately describe the spatial distribution of the electrons that spill outside the medium's ionic surface, and to account for all surfaces (free and bound electrons alike) in and around the transition region indicated by the dashed line. In addition, nonlinear optical phenomena due to anharmonic spring behavior (Fig. 1 (c)) is necessarily confined to the volume defined by the surface nuclei, i.e. to the left of the dashed line in Fig. 1 (b). The free electron gas spilling out of the surface beyond the nuclei shields nonlinear third order effects arising from bound electrons. Measurements of this effect are referred to as the Metal-Induced Gap States or MIGS [42].

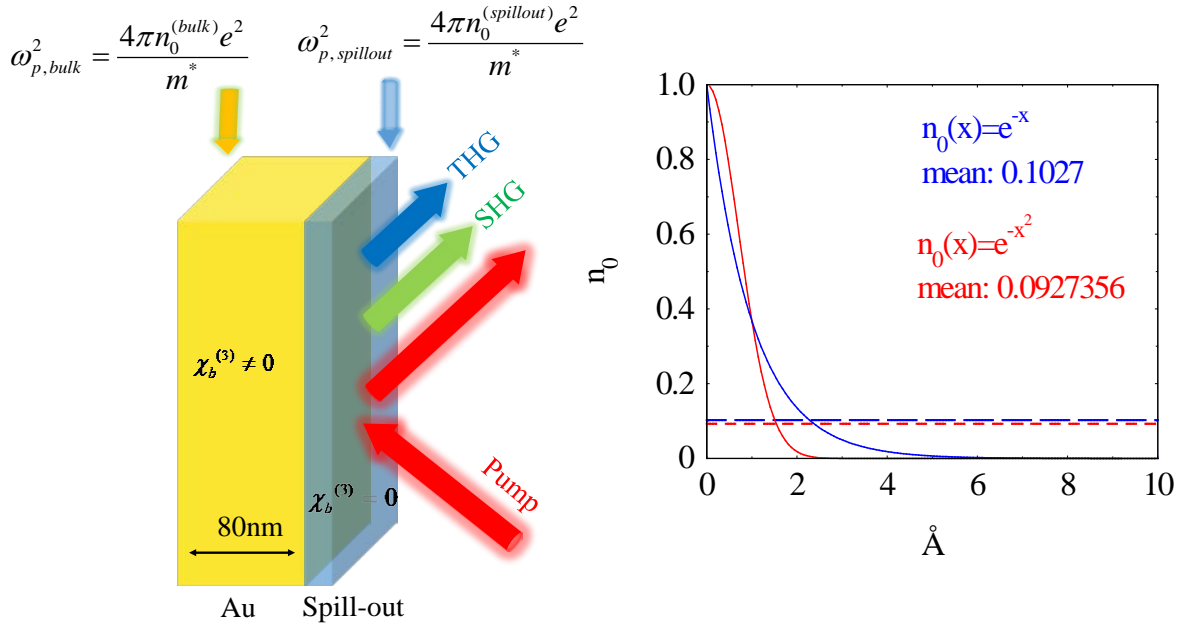

**Figure 2**: Right Panel: depiction of two types of exponential decay of the electron cloud that covers the metal surface. Both decays yield a mean density approximately 10% of the value at the hard, ionic surface, i.e. the dashed line in Fig. 1. The main difference between the two density profiles is the spatial extension into vacuum, i.e. 2Å-5Å. Left Panel: Once the average density and spatial extension into vacuum have been chosen, the electromagnetic problem is solved by introducing an external layer that contains only free charges, and thus two surfaces, resulting in a discontinuous

free charge density and a screened internal medium composed of free and bound charges. The bulk, third order nonlinearity is assumed to originate only in bound charges, which are described as collections of Lorentz oscillators.

Calculations have shown that for real metals the charge density decays exponentially with distance from the ionic surface [43]. Therefore, we adopt a similar model, at first assuming some type of exponential decay of the free charge density from the surface (Fig. 2, right panel), and subsequently assigning average value and thickness to an external charge density that in our modified scheme forms a single, uniform layer composed only of free charges (Fig. 2, left panel). Fig. 2 shows the modified configuration of Fig. 1.

The local dielectric constant of the free electron layer is Drude-like and given by:

$$\varepsilon(\omega)_{spillout} = 1 - \frac{\omega_{pf,spillout}^2}{\omega^2 + i\gamma_f \omega}, \quad (1)$$

$\omega_{pf,spillout} = \sqrt{\frac{4\pi n_{spill-out} e^2}{m_0}}$ is the plasma frequency and $\gamma_f$ is the related damping coefficient. The local dielectric constant of the interior bulk section also contains a Drude portion that describes free electrons, and at least two Lorentz oscillator contributions that allows one to model a more accurate medium response for wavelengths down to approximately 200nm, that take into account the contribution to the dielectric constant by d-shell electrons, as follows:

$$\varepsilon_{bulk}(\omega) = 1 - \frac{\omega_{pf,bulk}^2}{\omega^2 + i\gamma_f \omega} - \frac{\omega_{p1}^2}{\omega^2 - \omega_{01}^2 + i\gamma_{01}\omega} - \frac{\omega_{p2}^2}{\omega^2 - \omega_{02}^2 + i\gamma_{02}\omega}. \quad (2)$$

$\omega_{p1,2}$ are the bound electrons' plasma frequencies and $\gamma_{01,2}$ the related damping coefficients. The ionic surface represented by the dashed line in Fig.1 demarcates both free and bound electron discontinuities and contains surface nonlinearities of its own [24, 25]. For simplicity we assume that free electrons found inside the volume have identical damping coefficients as free electrons in the spill-out layer, but may have different densities/plasma frequencies. In addition to bound electrons, the total *linear* dielectric function is modified dynamically by a second order spatial derivative of the free electron polarization that describes electron gas pressure in the two relevant regions of space depicted in Fig. 2. For now we neglect nonlocal effects due to viscosity [39]. Most of what occurs at the surface and the evolution of the generated signals may, under the right circumstances, be determined entirely by the density of the thin, external layer of free charges. The dynamical equation of motion that describes harmonic generation from the free electron gas portions, modified to account for a discontinuous charge density (i.e. a spatial derivative,) nonlocal pressure effects, magnetic contributions and convection may be written as follows [44]:

$$\ddot{\mathbf{P}}_f + \gamma_f \dot{\mathbf{P}}_f = \frac{n_0 e^2}{m^*}\left(\frac{\lambda_0}{c}\right)^2 \mathbf{E} - \frac{e\lambda_0}{m^* c^2}(\nabla \cdot \mathbf{P}_f)\mathbf{E} + \frac{e\lambda_0}{m^* c^2}\dot{\mathbf{P}}_f \times \mathbf{H}$$
$$- \frac{1}{n_0 e \lambda_0}\left[(\nabla \cdot \dot{\mathbf{P}}_f)\dot{\mathbf{P}}_f + (\dot{\mathbf{P}}_f \cdot \nabla)\dot{\mathbf{P}}_f + n_0 \dot{\mathbf{P}}_f(\dot{\mathbf{P}}_f \cdot \nabla)\left(\frac{1}{n_0}\right)\right] + \frac{5 E_F}{3 m^* c^2}\nabla(\nabla \cdot \mathbf{P}_f) \quad (3)$$

where $\mathbf{P}_f$ is the free electron polarization; $\mathbf{E}$ and $\mathbf{H}$ are the propagating electric and magnetic fields, respectively; $m^* = m_e$ is the free electron's effective mass; $\gamma_f = 10^{14}$ rad/sec is the damping coefficient; $n_0 = 5.8 \times 10^{22}/\text{cm}^3$ is the background charge density with no applied field; $E_F = \frac{\hbar^2}{2m^*}(3\pi^2 n_0)^{2/3} \approx 5\text{eV}$ is the Fermi level of gold; $c$ is the speed of light in vacuum. We note that the description of SHG and THG requires that Eq.(3) split into three coupled equations, each describing one of the harmonic fields [24]. The equation is scaled with respect to dimensionless time, longitudinal and transverse coordinates (2-D), $\tau = ct/\lambda_0$, $\xi = z/\lambda_0$, $\tilde{y} = y/\lambda_0$, respectively, where $\lambda_0 = 1\mu m$ is chosen as a convenient reference wavelength. A *discontinuous free charge density* between the external and internal free electron distributions is equivalent to a metal/metal interface, and is described by the nonlinear term that appears inside the bracketed expression on the right hand side, i.e. $n_0 \dot{\mathbf{P}}_f (\dot{\mathbf{P}}_f \cdot \nabla)(1/n_0)$. Eq.(3) represents a simple Drude model when $(n_0 e^2 \lambda_0^2 / m^* c^2)\mathbf{E}$ is the only driving term, augmented by a number of linear and nonlinear source terms as follows: the magnetic Lorentz force, $(e\lambda_0/m^* c^2)\dot{\mathbf{P}}_f \times \mathbf{H}$; a Coulomb term, $-(e\lambda_0/m^* c^2)\mathbf{E}(\nabla \cdot \mathbf{P}_f)$ describes redistribution of free charges at and near each boundary, according to the strength of the derivatives; convective terms, $\left[(\nabla \cdot \dot{\mathbf{P}}_f)\dot{\mathbf{P}}_f + (\dot{\mathbf{P}}_f \cdot \nabla)\dot{\mathbf{P}}_f + n_0 \dot{\mathbf{P}}_f(\dot{\mathbf{P}}_f \cdot \nabla)(1/n_0)\right]$; and a linear electron gas pressure term proportional to $\nabla(\nabla \cdot \mathbf{P}_f)$ that leads to a k-dependent dielectric constant. Similarly to Eq. (3), each species of bound electrons is described by a nonlinear oscillator equation of the following type [24, 25]:

$$\ddot{\mathbf{P}}_1 + \gamma_{01}\dot{\mathbf{P}}_1 + \omega_{01}^2 \mathbf{P}_1 - b_1(\mathbf{P}_1 \cdot \mathbf{P}_1)\mathbf{P}_1 = \frac{n_{01} e^2 \lambda_0^2}{m_{b1}^* c^2}\mathbf{E} + \frac{e\lambda_0}{m_{b1}^* c^2}\dot{\mathbf{P}}_1 \times \mathbf{H}. \quad (4)$$

A similar equation describes the second bound electron species. Eq.(4) needs further processing before surface contributions can be harnessed, and similarly to Eq.(3), each bound electron equation separates into three equations, each describing an harmonic of the field [24,25]. Bound electrons are characterized by effective mass $m_{b1}^*$; resonance angular frequency $\omega_{01}$; density $n_{01}$; damping $\gamma_{01}$; and third order nonlinear spring constant, $b_1$, which is generally proportional to $\chi^{(3)}$ and responsible for self-phase modulation and third harmonic generation (THG) [45]. The bound electron masses and densities are assumed to be the same as for free electrons, and the following coefficients: $\gamma_{b,1} = \gamma_{b,2} = 2.2 \times 10^{15}$ rad/s, $\omega_{0,1} = 16\pi \times 10^{14}$ rad/s for the first oscillator, and $\omega_{0,2} = 2.7\pi \times 10^{15}$ rad/s for the second oscillator. These oscillators, depicted in Fig. 1, directly generate a third harmonic signal. However, free electrons are also capable of generating a third harmonic signal via a cascaded process, i.e. sum-frequency up-conversion due to pump and second-harmonic mixing through any of the nonlinear source terms in Eq.(3). Ultimately, the material equations of motion yield a polarization that is the vectorial sum of each contribution, namely $\mathbf{P}_{Total} = \mathbf{P}_f + \mathbf{P}_1 + \mathbf{P}_2 + ...$, which in turn is inserted into Maxwell's equations to solve for the dynamics. Therefore, unlike most models, which routinely exclude valence electrons from the full dynamics, neglect the ionic surface, and assume the pump remains undepleted, our approach allows for self-phase modulation, pump depletion, band shifts, nonlocal effects, free and bound charge interfaces, and for linear and nonlinear, d-shell electron contributions to the dielectric function.

According to our prescription in Fig.2, a gold mirror may thus be thought of as a two-layer system characterized by a metal/metal interface: a free electron layer having a thickness between 2Å-5Å that covers a medium containing a mix of free and bound electrons. A similar free-electron layer should be considered on the right side of the mirror, but its effects are negligible for thick layers. The description then becomes a usual boundary value problem where the composition of individual layers and their thicknesses are chosen according to the quantization of atomic orbitals.

**Gold Mirror**

In reference [44] we presented the results of a study of SH and TH conversion efficiencies from a gold surface covered by a thin, Å-thick patina of free electrons of variable density, $n_{spill-out}$, that displays ENZ conditions. We demonstrated that: (i) reasonable conditions (densities) exist such that $\varepsilon(\omega)_{spillout}$, $\varepsilon(2\omega)_{spillout}$ and $\varepsilon(3\omega)_{spillout}$ approach near-zero values, thus triggering resonant conditions for the fields and correspondingly high conversion efficiencies [46]; (ii) if the free electron density is discontinuous, the term $n_0 \dot{\mathbf{P}}_f \left( \dot{\mathbf{P}}_f \bullet \nabla \right)(1/n_0)$ in Eq.(3) cannot be neglected as its presence can increase conversion efficiencies by several orders of magnitude compared to neglecting the spill-out effect; (iii) the thickness of the outer, free electron layer plays only a minor role on conversion efficiency. We also showed that under ENZ conditions, the intensity of the surface generated TH signal could overwhelm the TH signal originating in the bulk [44].

While it may be possible to engineer a surface charge density by technological means, it is perhaps more practical to exploit material dispersion to seek out resonant, ENZ conditions for a given density, an approach that we pursue below. For example, in Fig. 3 (a) we plot the wavelength for which the real part of the dielectric function of the external, free electron layer (Eq.1) becomes zero as a function of its density, normalized to the free electron density within the bulk. Each of the colored arrows in Fig. 3 (a) points to the free electron densities on the abscissa that connect to the wavelengths that correspond to the $\text{Re}(\varepsilon_{spill-out}) \approx 0$ conditions on the ordinate, i.e. approximately 400nm, 450nm and 525nm. In Fig. 3 (b) we plot the predicted SHG conversion efficiency vs. wavelength for these three fixed densities and 45⁰ incident angle, and compare with the SHG efficiency without an external free electron layer. TM-polarized, incident pulses are approximately 50fs in duration, with peak intensities of order 2GW/cm². The external free electron patina is assumed to be 0.25nm thick, although thickness seems to be unimportant [44], while the rest of the gold layer is 80nm thick. The figure suggests that for the plausible densities that we have chosen, a spectral analysis of the SH signal should reveal marked maxima at the ENZ conditions [46]. Alternatively, the presence of a peak in the SH spectrum similar to that reported in Fig. 3 could help determine the effective density of the free electron patina. Secondary maxima evolve at shorter wavelengths, most likely due to the ENZ conditions of the internal portions of the medium.

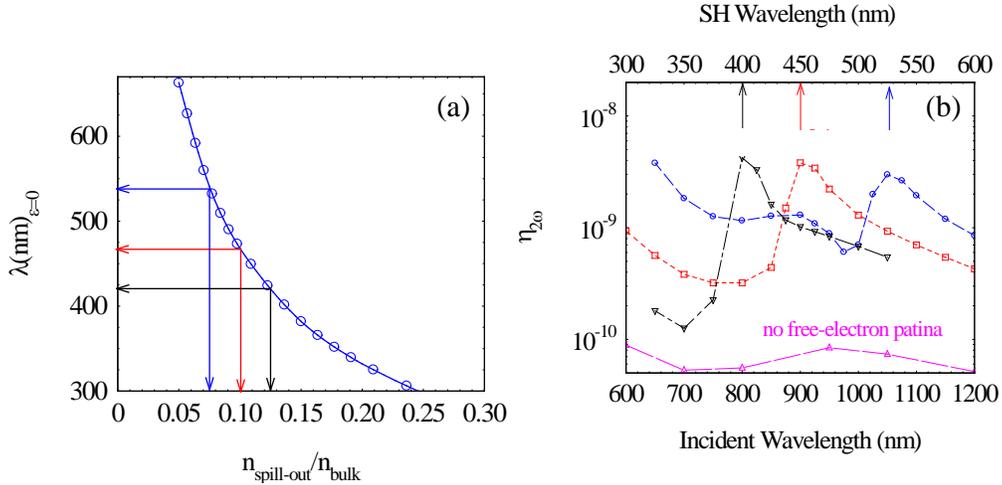

**Figure 3**: (a) Wavelength vs. density of external free electron layer where Re(ε)=0. Three possible densities are highlighted: $n_{spill-out} = 0.075 n_{bulk}$; $0.1 n_{bulk}$; $0.125 n_{bulk}$. These densities are close to the expected average density of the external free electron layer, as per the discussion surrounding Fig. 2. (b) SHG conversion efficiency vs. incident pump wavelength for each of the indicated densities, and for the case of no free electron patina. The SHG maxima occur where $\text{Re}[\varepsilon(2\omega)] \approx 0$. Additional maxima may occur at shorter wavelengths corresponding to zero-crossing of the real part of the internal portion dielectric constant, i.e. Eq.(2). The pump is incident a 45° and is TM polarized.

## ITO/Gold bilayer

In practical terms the situation described above, namely a discontinuous, free charge density, may be replicated by considering a thin layer of a free electron system like ITO deposited on a gold layer. The known charge density of commercially available ITO is approximately 100 times smaller compared to that of noble metals, i.e. $n_{ITO} \approx 5.8 \times 10^{20} / \text{cm}^3$, with an ENZ condition near 1246nm. These values may change depending on doping and annealing temperature. The model outlined above has been applied to study the nonlinear, high-gain context of nested plasmonic resonances [48], where field intensity is enhanced by the overlap of the intrinsic ENZ condition in an ITO particle placed inside the plasmonic resonance of a metallic nanoantenna.

The first structure that we consider is a 20nm ITO layer on a 80nm-thick gold layer. The geometrical configuration is similar to the gold mirror depicted in Fig.1, except that the outer, free electron layer now consists of ITO (Fig.4). For ITO, the free electron parameters are: $m^*_{ITO} = 0.5 m_e$, $\gamma_f = 2 \times 10^{13}$ rad/s and a corresponding $E_F \approx 1$ eV. These choices yield Fermi velocities that are similar for both gold and ITO, i.e. ~10⁶ m/s. The bound electron response in ITO (i.e., $\varepsilon_\infty$) is modelled with one Lorentz-type oscillator having the following coefficients: $m^*_{0b} = m_e$, $n_{0b} \approx 5.8 \times 10^{20} / \text{cm}^3$, $\gamma_b = 6\pi \times 10^{16}$ rad/s, $\omega_0 = 2\pi \times 10^{16}$ rad/s. For simplicity, the coefficient $b_b = \omega_0^2 / \left( n_{0b}^2 e^2 | \mathbf{r}_0^2 | \right)$ [48] and is chosen to be $b_b = 10^{37}$ m⁴/A² for both ITO and gold. In Fig.4 we plot SHG conversion efficiency vs. incident pump wavelength at fixed incident angle of 45º, to determine the importance of nonlocal terms and the additional nonlinear, convective source, $n_0 \dot{\mathbf{P}}_f (\dot{\mathbf{P}}_f \bullet \nabla)(1/n_0)$. The figure shows that both mechanisms introduce significant qualitative and quantitative differences in both transmitted and reflected signals, although the transmitted SH signal is more than three orders of magnitude weaker than the corresponding reflected signal due to the thickness of the gold layer.

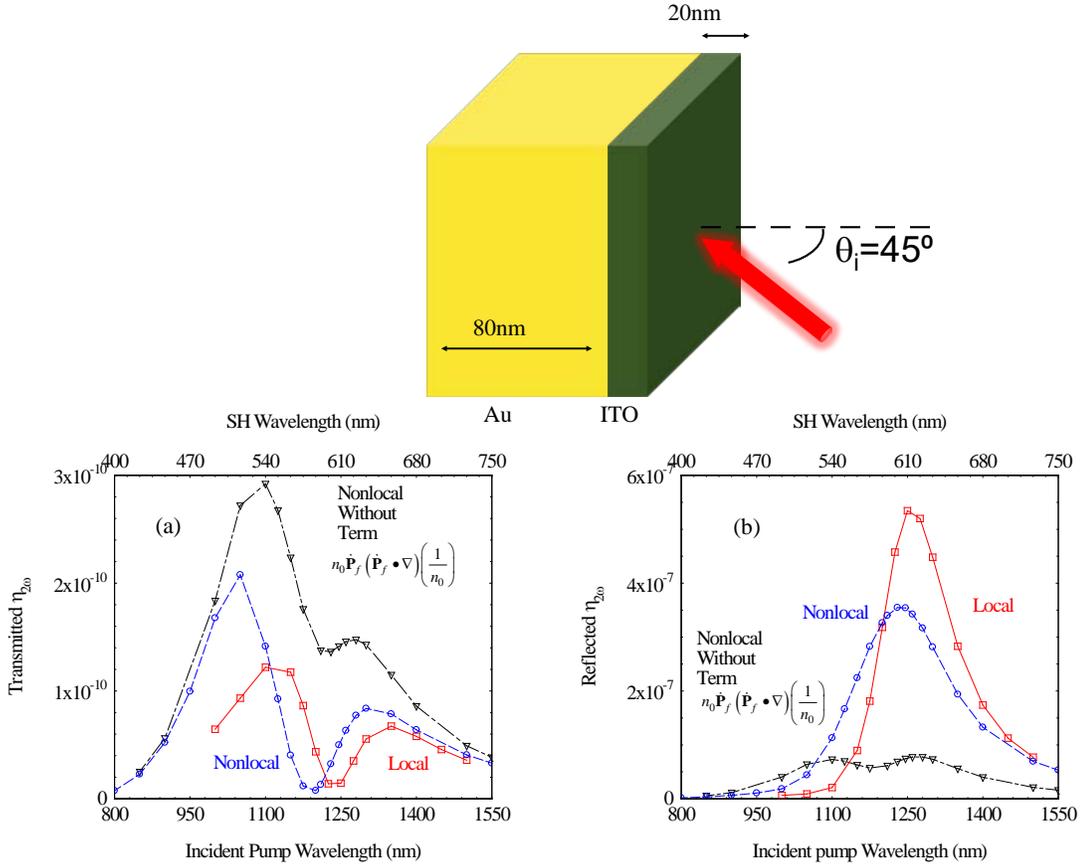

**Figure 4**: Top inset: geometrical depiction of the ITO(20nm)/Au(80nm) bilayer. (a) Transmitted and (b) reflected SHG efficiencies vs. incident pump wavelength. The angle of incidence is fixed at 45º. Both spectra display measurable shifts due to nonlocal effects. The inclusion of additional convective term due to the spatial discontinuity of the free electron density leads to substantial qualitative and quantitative differences for the reflected SHG.

As mentioned previously, there are two sources of THG: the bulk third order nonlinearity associated with the *b* coefficient in Eq.4, and the cascading process arising from surface and bulk terms in the free electron portions of both ITO and gold. In Fig.5 we plot the angular dependence of transmitted and reflected THG with ($b \neq 0$) and without ($b = 0$) a bulk third order nonlinear coefficient for the ITO/Gold bilayer depicted in Fig. 4. It is evident that most of the reflected signal is independent of *b*, arising mostly from the free electron component of ITO tied to the ENZ condition [44]. On the other hand, the peak of the transmitted component shifts by approximately 15° and is significantly influenced by the fact that the transmitted signal has to traverse the medium, although transmission is strongly abated by the thickness of the gold layer.

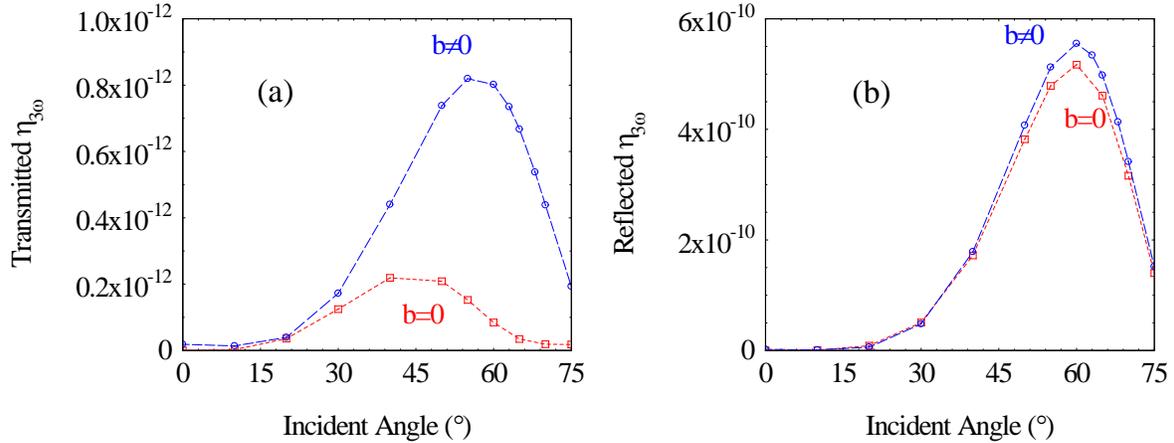

**Figure 5**: (a) Transmitted and (b) reflected THG efficiencies vs. incident angle for a pump wavelength of 1246nm, which coincides with the ENZ condition of ITO. The fact that the reflected component is barely affected when b=0 suggests that with these parameters the reflected TH signal originates mostly in the free electron (centrosymmetric) portions of the ITO. On the other hand, while the transmitted signal is orders of magnitude smaller than the reflected signal, the pulse crosses the sample before exciting, making transmittance more susceptible to bulk parameters.

We now examine the two-layer structure shown in Fig. 6, where ITO and gold layers have the same 20nm thickness, and light can be incident from either side. We focus on the ITO/Au transition region by neglecting any spill-out effect on either side. The limited thickness of gold insures that transmitted and reflected conversion efficiencies will be similar. In the figure we plot transmitted and reflected conversion efficiencies vs. incident angle for carrier wavelengths of 1246nm and 1064nm, and for right-to-left (RTL) and left-to-right (LTR) propagation. When the field is incident from the ITO side − Figs.6 (a) − the maximum reflected SH conversion efficiency is practically identical to the reflected conversion efficiency reported in Fig. 4 (b), where the gold layer is 80nm thick. This underscores the fact that the precise thickness of gold is unimportant, as the field is able to exploit the ENZ condition at 1246nm, leading to larger reflected SHG conversion efficiencies and a peak near 60⁰. In contrast, for RTL incidence, Fig. 6(b), the gold shields the ITO to the extent that the intensity inside the ITO is nearly one order of magnitude smaller compared to LTR incidence, shifting maxima closer to 70º, thus limiting conversion efficiencies with little or no influence from the ITO and mimicking a metal-only response [49].

In Fig. 6(c) and (d) we plot our predictions of THG conversion efficiencies vs. incident angle for LTR (6c) and RTL (6d) directions of propagation for the two-layer system shown in Fig.6.

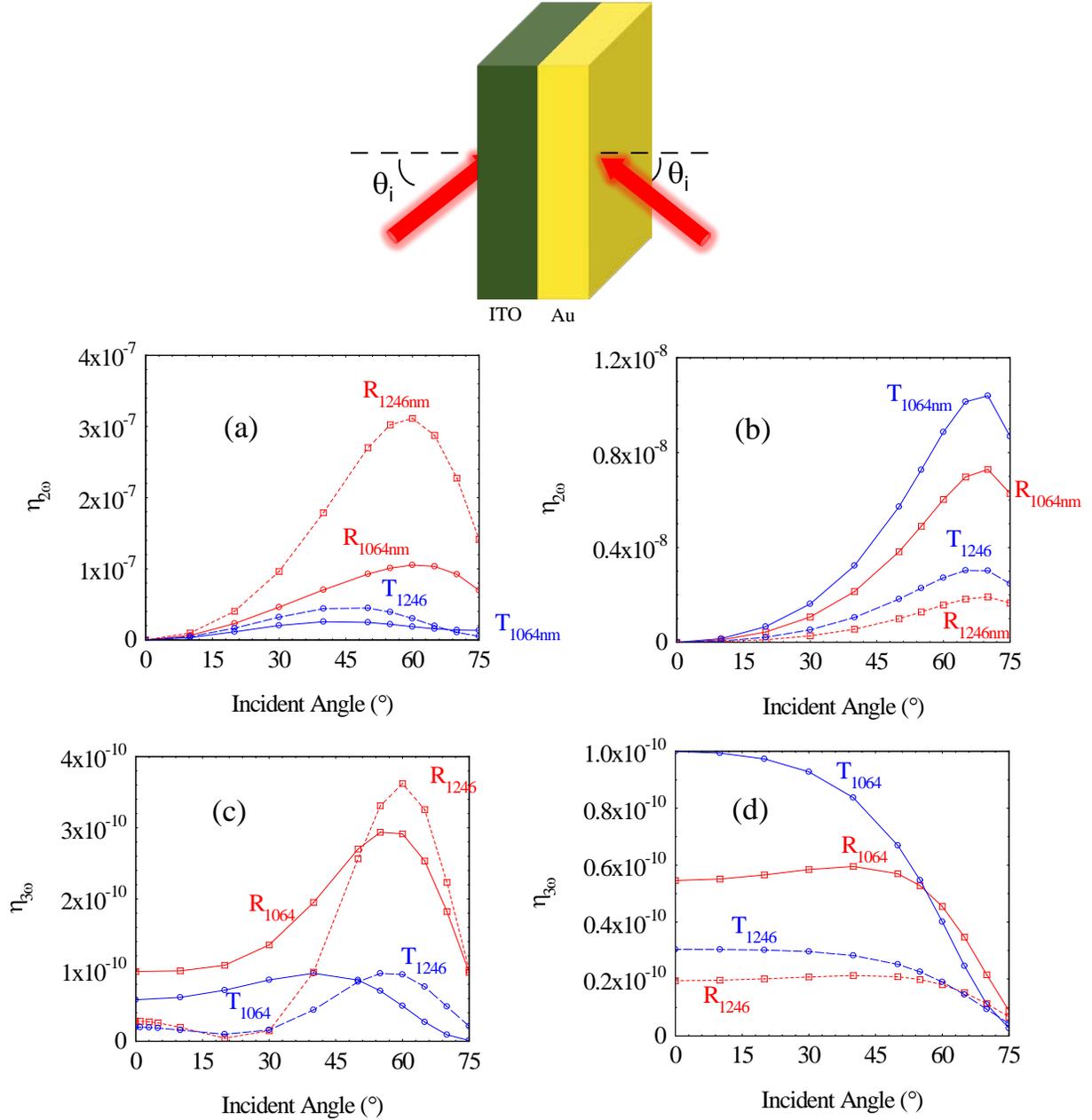
13

**Figure 6**: Reflected and transmitted SHG (a-b) and THG (c-d) conversion efficiencies for carrier wavelengths at 1246nm (ITO's ENZ condition) and 1064nm vs. incident angle for left-to-right (LTR) (a, c) and right-to-left (RTL) (b, d) incidence. The ITO's ENZ condition is best exploited for LTR incidence, while for RTL incidence, the gold layer shields the ITO reducing its effectiveness, and shifting the SHG peaks to larger angles. THG in the case of RTL (d) incidence resembles the results from a thin, metal-only layer, as reported in reference [48]. LTR incidence (c) is more interesting primarily because conversion efficiency displays minima near 20º at the ENZ condition.

The *b* coefficient has been chosen to have similar magnitudes in both metal and ITO sections. However, the gold presents a resonant $\chi_{3\omega}^{(3),Au}$ in the visible range, while the $\chi_{3\omega}^{(3),ITO}$ is designed to be resonant at much shorter wavelengths. Therefore, the bulk metal nonlinearity becomes more consequential for THG because the fields penetrate into both layers. The conversion efficiency profiles are thus strongly impacted by the direction of approach, and are more metal-like for the RTL direction of incidence [49]. Worthy of note are the THG minima that occur for both transmitted and reflected components near a 20º angle of incidence for LTR incidence, at the ENZ condition of ITO, which also appears for the ITO layer without gold backing. Obviously here we are not interested in pursuing any optimization of conversion efficiencies, but rather in validating the complex model that we use to point out the physical characteristics of harmonic generation from adjacent layers of free electron gas systems also as a function of direction of incidence. Finally, we note that our calculations include the term $n_0 \dot{\mathbf{P}}_f (\dot{\mathbf{P}}_f \bullet \nabla)(1/n_0)$, *whose magnitude is largest, but shielded,* for RTL propagation at the Au/ITO interface. For either direction of approach the magnitude of the derivative allows us for the moment to ignore the free electron patinas that extend into vacuum on either side of the stack.

## Summary

We have analyzed several examples of harmonic generation from interfacial regions that display free electron discontinuities. Our results suggests that it is possible to observe a nonlinear signature of the ENZ region over the surface of a simple metallic gold mirror. SHG and THG signals display maxima as the pump wavelength is tuned through the ENZ region. Additionally, the nonlinear signals have distinct shapes at fixed pump wavelength as the incident angle is changed. Issues relating to two- and multi-photon luminescence [48] could be dealt with by limiting pulse durations to under 100fs, and peak intensities under a few GW/cm$^2$. In order to overcome the intrinsic limitations of a gold surface, materials like ITO or CdO could be used as free electron systems that are known to display ENZ conditions of their own. Our results indeed suggest significant discrimination for LTR and RTL directions of propagation for thin layers, and significant impact of both nonlocal and free electron discontinuities on the dynamics.

## Acknowledgement

Financial support from U.S. Army RDECOM Acquisition Grant No. W911NF-15-1-0178 for JWH is gratefully acknowledged. JT and CC acknowledge financial support from RDECOM Grant W911NF-16-1-0563 from the International Technology Center-Atlantic.